\documentclass[12pt,preprint]{aastex}

\begin{document}

\title {OTHER KUIPER BELTS}

\author{M. Jura} 
\affil{Department of Physics and Astronomy, University of California,
    Los Angeles CA 90095-1562; jura@clotho.astro.ucla.edu}

\begin{abstract}

When a main sequence star evolves into a
red giant and its Kuiper Belt Object's (KBO's) reach a temperature of ${\sim}$170 K, the dust released during the rapid ice-sublimation of these cometary bodies 
may lead to a detectable infrared
excess at 25 ${\mu}$m, depending upon the  mass of the KBO's. Analysis of IRAS data for 66 first ascent red giants with 200 L$_{\odot}$ $<$ $L$ $<$ 300 L$_{\odot}$ within 150 pc of the Sun provides an   upper limit to the   mass in KBO's at 45 AU orbital radius that  is usually less than ${\sim}$0.1 M$_{\oplus}$.   
With improved infrared data, we may detect
systems of KBO's around first ascent red giants that are analogs to our Solar System's KBO's.

\end{abstract}

\section{INTRODUCTION}
The formation and evolution of large solids such as comets, asteroids and planets in astrophysical environments is of great interest.  Here, we investigate whether other stars possess Kuiper Belt Objects (KBO's) similar to
those found in the Solar System.

We assume that the hypothetical KBO's around other stars resemble those in our Solar System and have  a composition of ice and dust similar to that of comets (see Luu \& Jewitt 2002).  During the main sequence phase of the star's evolution, the KBO's quiescently remain in stable orbits.  However, when the star becomes a red giant,   
the KBO's  become sufficiently hot that ice sublimates and  previously embedded dust particles are ejected into the surroundings.   In $\S$2, we sketch our argument,  which is presented in full detail in $\S$3, that 
sufficient dust may be  released during this red giant phase that the star can display a detectable infrared excess.

 Stern, Shull \& Brandt (1990) and Ford \& Neufeld (2001) computed the fate of ice sublimated from comets that have orbital radii greater  than 100 AU when the host star evolves onto the Asymptotic Giant Branch and attains a luminosity near
10$^{4}$ L$_{\odot}$.  Now that the Kuiper Belt in the Solar System  with most objects lying near  45 AU from the Sun  is becoming better understood (Luu \& Jewitt 2002), it is possible to imagine the response of such a system to a star's first ascent up the red giant branch when its luminosity exceeds 100 L$_{\odot}$.  However, although progress is being made, there are still significant uncertainties about the Kuiper Belt in the Solar System.  Recent estimates of its  mass range  from 0.01 M$_{\oplus}$ (Bernstein et al. 2003) to 0.1 M$_{\oplus}$ (Luu \& Jewitt 2002).  Therefore, although we describe a procedure to investigate Kuiper Belt-like systems around other stars,  our own outer Solar System is so poorly understood that an exact comparison is not yet possible.

Two  current descriptions of the KBO's differ mainly but not exclusively in estimates of the numbers of objects with radii smaller than ${\sim}$200 km.    With the expectation that short period comets originate in  the Kuiper Belt,  Luu \& Jewitt (2002) propose a rapid rise in the numbers of such smaller KBO's.  In contrast to this expectation,  Bernstein et al. (2003) used the Hubble Space Telescope to search for KBO's as faint as 28.3 mag, corresponding to radii of ${\sim}$15 km, and found many fewer objects than
expected from the  model given by Luu \& Jewitt (2002).   However, with this observational result of Bernstein et al. (2002),
the origin of the short period comets becomes a mystery. Perhaps
there is a bimodal distribution of KBO's.  In this paper, given the
uncertainties in current knowledge of the outer Solar System, we consider models both 
with and without large numbers of KBO's smaller than 200 km in radius.

Most first ascent red giants do not show any evidence for circumstellar dust.  The minority of  first ascent red giants with infrared excesses (Judge, Jordan \& Rowan-Robinson 1987, Jura 1990, 1999, Kim, Zuckerman \& Silverstone 2001, Smith 1998, Zuckerman, Kim \& Liu 1995, Plets et al. 1997) do not seem to have  sublimating KBO's, since the characteristic inferred dust temperature is less than 70 K instead of 170 K, as we predict in $\S$3.   Here, we focus on the possibility of detecting relatively warm dust from disintegrating KBO's.

\section{OVERVIEW}
Because the details of our model are somewhat complicated, we first present
a schematic overview of our analysis, while the complete description is presented in $\S$3.  Using our own Kuiper Belt as
a prototype,  we assume that the KBO's orbit the host star near 45 AU.
At this orbital separation,  we find that most of the KBO material survives in the solid phase as long as the temperature is less than ${\sim}$150 K which occurs as long as the star is less luminous than ${\sim}$170 L$_{\odot}$.  However, once the star's luminosity exceeds this value, the KBO's become warm enough that relatively rapid ice sublimation occurs.  In the time interval, approximately characterized by the parameter $t_{RGB}$, during which 
the star increases in luminosity from ${\sim}$170 L$_{\odot}$ to ${\sim}$300 L$_{\odot}$,   most of the mass of the KBO's is lost
into their surroundings.    As the KBO's are sublimating, 
some dust particles are released that continue to orbit the star.  If sufficient mass is released from the KBO's, these orbiting dust grains can absorb enough
light to produce a detectable infrared excess.

 If $M_{KBO}(0)$ denotes the total initial mass of the KBO's and $f_{orbit}$ denotes the fraction of this mass which is ejected as dust particles that are large enough
to remain in orbit around the star and are not blown out of the system by radiation pressure, then the total amount of mass that can orbit the
star is $f_{orbit}M_{KBO}(0)$.   We assume that the dust 
particles  are  spheres of radius $b$ and density ${\rho}_{dust}$, and they orbit the star of luminosity, $L_{*}$, at distance, $D$. While this assumption is relaxed in $\S$3.7, here, for simplicity, we assume that $D$ does not change during
a dust particle's evolution.  Also,  while other factors are considered in $\S$3.7, here, we assume that the lifetime of the orbiting dust particles is controlled simply by
the Poynting-Robertson time, $t_{PR}$.   
Following Burns, Lamy \& Soter (1979), we write:
\begin{equation}
t_{PR} = \frac{4{\pi}b{\rho}_{dust}c^{2}D^{2}}{3L_{*}}
\end{equation}

Since the Poynting Robertson lifetime is typically shorter than the 
 the characteristic evolutionary time on the red giant branch, at any given moment during the star's evolutionary phase when grains are being released, the
fraction of orbiting material contributing to the infrared excess is $t_{PR}/t_{RGB}$. 
If the grains  have opacity, ${\chi}$ (cm$^{2}$ g$^{-1}$), and if the circumstellar dust cloud is optically thin, then the total luminosity of the circumstellar dust, $L_{excess}$ is scaled from the fractional dust coverage of an imaginary sphere of surface area 4${\pi}D^{2}$ by the relationship:
\begin{equation}
L_{excess}\;{\approx}\;\left(\frac{{\chi}f_{orbit}\,M_{KBO}(0)}{4{\pi}D^{2}}\right)\left(\frac{t_{PR}}{t_{RGB}}\right)\,L_{*}
\end{equation}
We use a simple model of the grain opacity where the geometric cross section (${\pi}b^{2}$) equals
the total absorption cross section.  In this case, since each dust grain has mass ($4{\pi}{\rho}_{dust}b^{3}/3$), we may write that:  
\begin{equation}
{\chi}\;=\;\frac{3}{4{\rho}_{dust}b}
\end{equation}
Combining equations (1)-(3), we find: 
\begin{equation}
L_{excess}\;{\approx}\;\frac{f_{orbit}\,M_{KBO}(0)\,c^{2}}{4\,t_{RGB}}
\end{equation}
 Expression (4)  captures the essence of our model.  The
infrared luminosity simply scales as the rate of production of dust grains which varies as $M_{KBO}(0)/t_{RGB}$.   If $f_{orbit}$ and $M_{KBO}(0)$ are sufficiently large, then the infrared excess may be detectable. Our estimate of L$_{excess}$, which is computed more exactly in $\S$3, is independent of the size of the KBO's as long as these objects are fully destroyed by
sublimation.

In $\S$3, we find characteristic values of $f_{orbit}$ ${\approx}$ 0.25 and
$t_{RGB}$ ${\approx}$ 14 Myr.  Therefore, if $M_{KBO}(0)$ ${\sim}$ 0.1 M$_{\oplus}$,
then from equation (4), $L_{excess}$ ${\approx}$ 0.02 L$_{\odot}$.   Because the KBO's mostly sublimate near 170 K,  
the spectrum of this  excess luminosity peaks at wavelengths near 25 ${\mu}$m.
In $\S$4, we show that a 170 K black body with a luminosity of 0.02 L$_{\odot}$   is just at the threshold of having been detected with IRAS data  for red giants with total luminosities less than 300 L$_{\odot}$. Consequently, the absence
of a measurable infrared excess at 25 ${\mu}$m around these red giants can be used to place an upper limit to the mass of KBO's closer than ${\sim}$50 AU of ${\sim}$0.1 M$_{\oplus}$.  In $\S$5, we discuss our results and in $\S$6, we present our conclusions.

\section{DETAILED MODEL}
We now elaborate about the schematic model described in $\S$2.
To compute the fate of the  KBO's  when a star
is evolving as a red giant, we write for the total mass loss rate from KBO's, ${\dot M_{KBO}}$, that:
\begin{equation}
{\dot M_{KBO}}\;=\;{\dot {\sigma}}_{KBO}(T)\,A_{KBO}(t)
\end{equation}
In equation (5), ${\dot {\sigma}}_{KBO}$ denotes the  mass loss rate per unit surface area of the KBO's as a function of temperature, $T$, while
$A_{KBO}(t)$ denotes the total surface area of these objects as a function of
time, $t$.  In $\S$3.3 and $\S$3.4, we describe calculations for ${\dot {\sigma}}_{KBO}$ and $A_{KBO}$ as functions of time or temperature, and
we  show in $\S$3.7 how  ${\dot M_{KBO}}$ can be used to estimate the infrared excess.    

\subsection{Initial Mass and Total Area of the KBO's}
    Here, we consider
two size distributions:
systems with initially many small (radius less than 200 km) KBO's, and Kuiper belts
with initially no KBO's smaller than 200 km in radius. 
\subsubsection{Model with initially many small KBO's}
Following Luu \& Jewitt (2002), we assume the KBO's are icy spheres
of radius, $a$, and follow  a broken power law for the size distribution, $n(a)\,da$. For $a$ $>$ $a_{break}$, 
\begin{equation}
n(a)\,da\;=\;n_{0}\,\left(\frac{a_{break}}{a}\right)^{4}\,da
\end{equation}
while for $a$ $<$ $a_{break}$,
\begin{equation}
n(a)\,da\;=\;n_{0}\,\left(\frac{a_{break}}{a}\right)^{3.5}\,da
\end{equation}
From Luu \& Jewitt (2002), we take  $a_{break}$ = 1 km and a maximum size, $a_{max}$, of about 1000 km.  The distribution in equation (7) for the  KBO's smaller than 1 km is derived from theoretical models
(Kenyon 2002) and is consistent with upper limits to the contribution of KBO's to the  zodiacal light and to distortions of the microwave
background (Kenyon \& Windhorst 2001, Teplitz et al. 1999). 

Before a star becomes a red giant, the initial mass of the larger KBO's, $M_{large}(0)$, is:
\begin{equation}
M_{large}(0)\;=\;{\int}_{a_{break}}^{a_{max}}n_{0}\,\left(\frac{a_{break}}{a}\right)^{4}\,\left(\frac{4{\pi}{\rho}_{KBO}a^{3}}{3}\right)\,da\;=\;\frac{4{\pi}}{3}{\rho}_{KBO}\,n_{0}\,a_{break}^{4}\,ln\left(\frac{a_{max}}{a_{break}}\right)
\end{equation}
where   ${\rho}_{KBO}$   denotes the density of the KBO's. Following Greenberg (1998) and Kenyon (2002), we assume that the KBO's are composed mostly of ice with some additional refractory dust and that ${\rho}_{KBO}$ = 1.5 g cm$^{-3}$.
From the measured size distribution of KBO's (Luu \& Jewitt 2002), we  adopt $M_{large}(0)$ = 0.1 M$_{\oplus}$ as a first approximation to the mass of KBO's.  Luu \& Jewitt (2002) 
use  ${\rho}_{KBO}$ = 1.0 g cm$^{-3}$ and  a slightly smaller
total mass of 0.08 M$_{\oplus}$.
  
The total initial surface area of
the large KBO's, $A_{large}(0)$, is
\begin{equation}
A_{large}(0)\;=\;{\int}_{a_{break}}^{a_{max}}n_{0}\,\left(\frac{a_{break}}{a}\right)^{4}\,(4\,{\pi}a^{2})\,da\;{\approx}\;4{\pi}n_{0}a_{break}^{3}\;{\approx}\;\frac{3M_{large}(0)}{{\rho}_{KBO}\,a_{break}\,ln\left(\frac{a_{max}}{a_{break}}\right)}
\end{equation}
For the parameters we adopted  above, $A_{large}(0)$ = 1.7 ${\times}$ 10$^{21}$ cm$^{2}$.

We can also compute the total mass contained in the KBO's with $a$ smaller than 1 km.  If $a_{min}$ denotes the smallest KBO, then:
\begin{equation}
M_{small}(0)\;=\;{\int}_{a_{min}}^{a_{break}}n_{0}\,\left(\frac{a_{break}}{a}\right)^{3.5}\,\left(\frac{4{\pi}{\rho}_{KBO}a^{3}}{3}\right)\,da\;{\approx}\;\frac{8{\pi}}{3}{\rho}_{KBO}\,n_{0}\,a_{break}^{4}
\end{equation}
As long as $a_{min}$ $<<$ $a_{break}$, then evaluation of the integral in equation (10) shows that the mass of the small KBO's is essentially independent of $a_{min}$.  
From equations (8) and (10), we find that:

\begin{equation}
M_{small}(0)\;=\;\frac{2\,M_{large}(0)}{ ln\left(\frac{a_{max}}{a_{break}}\right)}
\end{equation}  
Therefore, with our adopted values of $a_{max}$ and $a_{break}$, we find that  M$_{small}(0)$ ${\approx}$ 0.3 M$_{large}(0)$; most of the mass of the KBO's is contained in the larger objects.   

\subsubsection{Model with initially only large KBO's}

As a second model, we adopt a simplified version of the results of Bernstein et al. (2003).   In particular, we assume that the size distribution of KBO's follows equation (6), but there are no KBO's smaller than 
$a_{break}$ with   $a_{break}$  = 200 km instead of 1 km.  We denote the initial mass of the KBO's in this
system as $M_{big}(0)$.  If the numbers of objects larger than 200 km in radius is the same as in the model described in $\S$3.1.1, then $M_{big}(0)$ ${\approx}$ 0.02 M$_{\oplus}$.  Although this model is representative of the results of Bernstein et al. (2003), 
it does not agree exactly with their description of the Kuiper Belt which involves either a ``rolling power law" fit or a ``double power law" fit to the size distribution of the KBO's.   One source of uncertainty is that  even for
KBO's larger than 200 km in radius, there
is some difference between the descriptions of Luu \& Jewitt (2002) and Bernstein et al. (2003) which we cannot resolve in this paper. In any case, by evaluating the integral in equation (9) with the new lower limit to the size of 200 km, we find
that the initial area of the KBO's in this model, $A_{big}(0)$, is 8.7 ${\times}$ 10$^{18}$ cm$^{2}$.  
   
\subsection{Red Giant Branch Evolution}

The fate of a KBO is controlled by the host star's evolution.  For simplicity,
we  parameterize the detailed calculations  by 
Girardi et al. (2000) for a star's luminosity, $L_{*}$, on the red giant
 branch as:
 \begin{equation}
L_{*}(t)\;{\approx}\;L_{0}\, e^{t/t_{RGB}}
\end{equation}
For stars of 1 M$_{\odot}$, we  fit  equation (12) with $t_{RGB}$ = 1.4 ${\times}$ 10$^{7}$ yr, $L_{0}$ = 87 L$_{\odot}$ and a starting time ($t$ = 0) of 1.220 ${\times}$ 10$^{10}$ year after the star's birth. For stars with 1.5 M$_{\odot}$, we employ the same value of  $t_{RGB}$ and adopt a starting time of 2.7742 ${\times}$ 10$^{9}$ yr and $L_{0}$ = 91 L$_{\odot}$.  In Figure 1 we show a  comparison of the
prediction from equation (12) using $t_{RGB}$ = 1.4 ${\times}$ 10$^{7}$ yr with the detailed calculations by Girardi et al. (2000) for  stars of 1 M$_{\odot}$ and 1.5 M$_{\odot}$.  
For luminosities less than 1000 L$_{\odot}$, the agreement is mostly better than 20\%.
Although equation (12) is an imperfect match to the detailed calculations, it does illustrate that the growth of luminosity with
time on the red giant branch for a star of ${\sim}$ 150 L$_{\odot}$ is very approximately exponential with an e-folding time of ${\sim}$ 1.4 ${\times}$ 10$^{7}$ yr.

\subsection{Ice Sublimation}

With the  description given in $\S$3.2 of the star's evolution, we can  
 estimate the rate of ice sublimation from an individual KBO.  We start with the expression:
\begin{equation}
\frac{d}{dt}\;\frac{4{\pi}{\rho}_{KBO}a^{3}}{3}\;=\;-4{\pi}a^{2}{\dot {\sigma}_{KBO}}
\end{equation}
Therefore, at time $t_{f}$, the total decrease in the radius of a KBO, ${\Delta}a$, is given by the following solution to expression (13):
\begin{equation}
{\Delta}a(t_{f})\;=\;{\int}_{0}^{t_{f}}\frac{{\dot {\sigma}_{KBO}}(T(t))}{{\rho}_{KBO}}\;dt
\end{equation}
As in equation (5), $T$ denotes the grain temperature.
Note that ${\Delta}a$ is independent of $a$.

To estimate ${\dot {\sigma}}_{KBO}$ we follow Ford \& Neufeld (2001) who
denote this quantity as ${\dot m}$.  
Converting from the units they employ and using their expression for the vapor pressure of ice, 
we write that
\begin{equation}
{\dot {\sigma}_{KBO}}\;=\;\,{\dot {\sigma}_{0}}\,T^{-1/2}\,e^{-T_{subl}/T}
\end{equation}
where ${\dot {\sigma}_{0}}$ = 3.8 ${\times}$ 10$^{8}$ g cm$^{-2}$ s$^{-1}$ $K^{1/2}$ and $T_{subl}$ = 5530 K.  In  expression (15), the chemical binding energy of H$_{2}$O molecules onto the solid ice is $k_{B}T_{subl}$ where
$k_{B}$ is Boltzmann's constant.   

To compute ${\Delta}a$ from equation (14), we need to find the temperature of a KBO as a function of time.   
If we ignore  the difference between the day an night side of the KBO, assume unit emissivity and an albedo of 0, close to  0.07 which may be a typical for a KBO's albedo (Luu \& Jewitt 2002), then for a KBO at distance, $D_{init}$, from the star:
\begin{equation}
T(t)\;=\;\left(\frac{L_{*}(t)}{16{\pi}D_{init}^{2}{\sigma}_{SB}}\right)^{1/4}
\end{equation}
where ${\sigma}_{SB}$ is the Stephan-Boltzmann constant. In equation (16), we ignore cooling by sublimation. At 170 K, the
characteristic temperature when most of the ice has been sublimated, it can be shown from the results in Ford \& Neufeld (2001) that cooling by ice sublimation is  less than 7\% of the cooling
by radiation.

From equations (12) and (16),  we  write that
\begin{equation}
dt\;=\;4\,t_{RGB}\frac{dT}{T}
\end{equation}
Using equations (15) and (17), we  re-write equation (14) to find:
\begin{equation}
{\Delta}a(T_{f})\;=\;4\,\frac{t_{RGB}\,{\dot {\sigma}_{0}}}{{\rho}_{KBO}}{\int}_{0}^{T_{f}}T^{-3/2}\,e^{-T_{subl}/T}\,dT
\end{equation}
where $T_{f}$ is the KBO's temperature at time $t_{f}$.  
With the substitution that $u^{2}$ = $T_{subl}/T$,  the integral in equation (18) can be re-expressed as a complementary error function to give:
\begin{equation}
 {\Delta}a(T_{f})\;=\;8\,\frac{t_{RGB}\,{\dot {\sigma}_{0}}}{{\rho}_{KBO}\,T_{subl}^{1/2}}{\int}_{u_{f}}^{\infty}e^{-u^{2}}\,du
\end{equation}
When $u_{f}$ $>$ 1 which is equivalent to our case where $T_{subl}$ $>>$ $T$, then (see, for example, Abramowitz \& Stegun 1965)
\begin{equation}
{\int}_{u_{f}}^{\infty}e^{-u^{2}}\,du\;{\approx}\;\frac{1}{2}\frac{e^{-u_{f}^{2}}}{u_{f}}
\end{equation}
Therefore:
\begin{equation}
{\Delta}a(T_{f})\;{\approx}\;4\;\frac{t_{RGB}{\dot {\sigma}_{0}}}{{\rho}_{KBO}}\;\frac{T_{f}^{1/2}}{T_{subl}}\;e^{-T_{subl}/T_{f}}
\end{equation}
Equation (21) describes the shrinkage of  KBO's as a function of their temperature.    
Numerically,  for 
$T_{f}$ = 150 K, 160 K, and 171 K, then ${\Delta}a$ = 1 km, 10 km, and 100 km, respectively.  These temperatures are achieved by KBO's at 45 AU when the red giant's luminosity is 170 L$_{\odot}$, 220 L$_{\odot}$,  and 290 L$_{\odot}$, respectively.   Complete destruction of the KBO's is achieved when ${\Delta}a$  = 1000 km which occurs for KBO's at 45 AU, when the star's luminosity is ${\sim}$400 L$_{\odot}$.   In the model
with initially  many small KBO's  at 45 AU orbital radius described in $\S$3.1.1, by the time the temperature attains ${\sim}$170 K,  all the KBO's smaller than 100 km are destroyed and somewhat more than half of the total initial mass of KBO's has been sublimated.  Therefore, we adopt 170 K as a
characteristic reference temperature for production of dust from KBO's.   

\subsection{Evolution of the Total Area of the KBO's}

Using our calculation for  the evolution of the size of a single KBO, we now consider the evolution of the surface area of the entire ensemble to use in equation (5).     
Each KBO of initial radius $a$ evolves into
a KBO with radius $a'$ where
\begin{equation}
a'\;=\;a\,-{\Delta}a(t)
\end{equation}

\subsubsection{Model with initially many small KBO's}

For the large KBO's, we use equation (6) to find that their size distribution, $n(a')\,da'$, is:
\begin{equation}
n(a')da'\;=\;n_{0}\left(\frac{a_{break}}{a'\,+\,{\Delta}a}\right)^{4}\,da'
\end{equation}
To evaluate the total area of the KBO's with time, there are  two regimes to consider.  If
 ${\Delta}a$ $<$ $a_{break}$,  we write that
\begin{equation}
A_{large}(t)\;=\;{\int}^{a_{max}-{\Delta}a}_{a_{break}-{\Delta}a}(4\,{\pi}a'^{2})\,n_{0}\left(\frac{a_{break}}{a'\,+\,{\Delta}a}\right)^{4}\,da'
\end{equation}
Using  equation (9), we evaluate expression (24)  to find in the limit where $a_{max}$ $>>$ $a_{break}$ that: 
\begin{equation}
A_{large}(t)\;=\;A_{large}(0)\left(1\,-{\delta}\,+\,\frac{{\delta}^{2}}{3}\right)
\end{equation}
where ${\delta}$ = ${{\Delta}a}/{a_{break}}$ and $A_{large}(0)$.    
From equation (25), we see that the
total surface area of the large KBO's remains nearly constant  until ${\Delta}a$ approaches $a_{break}$.  Note that equation (25) is valid only when ${\delta}$ $<$ 1.   
 
When ${\Delta}a$ $>$ $a_{break}$  we must use a different lower bound in the
integral and we 
write that
\begin{equation}
A_{large}(t)\;=\;{\int}_{0}^{a_{max}-{\Delta}a}(4\,{\pi}a'^{2})\,n_{0}\left(\frac{a_{break}}{a'\,+\,{\Delta}a}\right)^{4}\,da'
\end{equation}
We evaluate equation (26) to find in the limit where $a_{max}$ $>>$ ${\Delta}a$ that:
\begin{equation}
A_{large}(t)\;{\approx}\;A_{large}(0)\frac{a_{break}}{3{\Delta}a}\;=\;\frac{A_{large}(0)}{3\,{\delta}}
\end{equation}
Equation (27) is valid when ${\delta}$ $>$ 1.  
In this regime,  the total area of the large KBO's  decreases as ${\Delta}a$ increases with time.  As ${\Delta}a$ 
approaches $a_{max}$, equation (27) overestimates $A_{large}(t)$.  However,  as long as ${\Delta}a$ $<$ 0.5 $a_{max}$, then we can find from equation (26) that equation (27) is accurate to better than a factor of two.

In the above paragraph, we described the evolution with time of the total surface
area of the KBO's initially larger than $a_{break}$.  In this paragraph, we describe the
evolution of the surface area of the smaller KBO's.    As long as ${\Delta}a$ $>$ $a_{min}$, we can write from equation (7) that:
\begin{equation}
A_{small}(t)\;=\;{\int}_{0}^{a_{break}-{\Delta}a}(4\,{\pi}a'^{2})\,n_{0}\left(\frac{a_{break}}{a'\,+\,{\Delta}a}\right)^{3.5}\,da'  
\end{equation} 
Expression (28)  can be conveniently evaluated  to give:
\begin{equation}
A_{small}(t)\;=\;\frac{3\,M_{small}(0)}{2\,{\rho}_{KBO}\,(a_{break}{\Delta}a)^{1/2}}\,\left(2\,(1\,-\,{\delta}^{1/2})\,-\frac{4}{3}(1\,-\,{\delta}^{3/2})\,+\,\frac{2}{5}(1\,-\,{\delta}^{5/2})\right)
\end{equation}
Note that $A_{small}$ approaches zero as ${\Delta}$a approaches $a_{break}$,  or, equivalently, as ${\delta}$ approaches unity.  

\subsubsection{Model with initially only large KBO's}

To calculate $A_{big}(t)$  we simply substitute for $A_{large}(t)$ in the  equations in  $\S$3.4.1  with 
 $a_{break}$ = 200 km instead of 1 km. When ${\Delta}a$ $<$ $a_{break}$, we use equation (25), and  when ${\Delta}a$ $>$ $a_{break}$, we use
equation (27).     

\subsection {Total KBO Mass Loss Rate}
Armed with the KBO temperature and total surface area as functions of time, we can evaluate equation (5) to find the total rate at which
KBO's loss mass into their surroundings. It is convenient to present 
${\dot M_{KBO}}$ as a function of the host star's luminosity, since that is a measurable quantity.    The star's
evolution is given by equation (12), and the sublimation rate from the KBO's is given by equation (15).  
\subsubsection{Model with initially many small KBO's}
 
For systems with many small KBO's, we consider two regimes for ${\dot M_{KBO}}$ as a function of the star's luminosity.  First,   when ${\Delta}a$ $>$ $a_{break}$, $A_{large}(t)$ from equation (27) gives the total surface area of the KBO's.  In this case, ${\dot M_{KBO}}$ reaches a saturation value and is nearly independent of $L_{*}$.  
At saturation, 
   we find
from equations (5), (9), (15), (21) and (27) that
\begin{equation}
{\dot M_{KBO}}(T)\;=\;\left(\frac{1}{4\,ln\left(\frac{a_{max}}{a_{break}}\right)}\right)\left(\frac{T_{Subl}}{T}\right)\left(\frac{M_{large}(0)}{t_{RGB}}\right)
\end{equation}
Equation (30) shows that when  when ${\Delta}a$ $>$ $a_{break}$, the sublimation rate from the KBO's depends mainly upon the initial mass of the large KBO's and the time scale of the star's evolution on
the red giant branch.   In this phase, the rapid rise in the sublimation rate with temperature given in equation (15) is roughly balanced by the decrease in the total area of KBO's given in equation (27).    When $T$ = 170 K, we find from equation (30) that ${\dot M_{KBO}}$ ${\approx}$ 1.2 $M_{large}(0)/t_{RGB}$, or 1.6 ${\times}$ 10$^{12}$ g s$^{-1}$.

A second regime to consider is when ${\Delta}a$ $<$ $a_{break}$.  In this
case, the total surface area of the KBO's is given by the sum of $A_{large}(t)$
and $A_{small}(t)$ from equations (25) and (29). In this regime, ${\dot M_{KBO}}$ starts low and rises rapidly with the star's luminosity until saturation is achieved. 
 
Graphical results for ${\dot M_{KBO}}$ vs. luminosity  and presented in Figure 2 for a  model with $D_{init}$ = 45 AU and $M_{large}(0)$ =  0.1 M$_{\oplus}$.  For comparison, we also show the results for models with $D_{init}$ of 40 AU or 50 AU.  In all three cases, we see that ${\dot M_{KBO}}$
approaches the saturation rate predicted by equation (30) for sufficiently large value of the star's luminosity.

\subsubsection{Model with initially only large KBO's}

If initially there are only large KBO's, then for much of the time during the star's
evolution  the total surface area of the KBO's is approximately constant.  As long as ${\Delta}a$ $<$ 200 km, we may write from equations (5). (15) and (25) that: 
\begin{equation}
{\dot M_{KBO}}(T)\;=\;A_{big}(0)\left(1\,-{\delta}\,+\,\frac{{\delta}^{2}}{3}\right)\,{\dot {\sigma}_{0}}\,T^{-1/2}\,e^{-T_{subl}/T}  
\end{equation}
Graphical results for ${\dot M_{KBO}}$ vs. luminosity from  equations (30) and (31) 
are shown in Figure 2 for the case where $D_{init}$ = 45 AU. We see that for stellar luminosities less than ${\sim}$ 300 L$_{\odot}$, the dust production rate is relatively  small in this regime where initially there are only large KBO's, but as the luminosity increases, the saturation value of ${\dot M_{KBO}}$ is attained.   
 
\subsection{Dust Production}

In $\S$3.5, we estimated the rate at which ice sublimates from the KBO's; here, we estimate the fractional mass of the KBO's released in dust  that continues to orbit the star, $f_{orbit}$.  It is this orbiting dust which controls the infrared excess of the system.

 It is plausible that half of the total mass of the  KBO's is dust  (Luu \& Jewitt 2002), and $f_{orbit}$ could be as large as 0.5.  However, we expect that $f_{orbit}$ is less than 0.5 since dust grains that are smaller than a critical radius, $b_{crit}$,  do not orbit the star of mass, $M_{*}$, and instead are driven
out  the system by radiation pressure.  The critical size for being gravitationally bound is (Artymowicz 1988):
\begin{equation}
b_{crit}\;=\;\frac{3L_{*}}{16{\pi}{\rho}_{dust}GM_{*}c}
\end{equation}
If the dust is composed of silicates with a density, ${\rho}_{dust}$, of 3 g cm$^{-3}$, then for a 1 M$_{\odot}$ star of 300 L$_{\odot}$, the maximum  luminosity that we consider,  $b_{crit}$ = 57 ${\mu}$m.

We assume that  the  size distribution of  dust  ejected from KBO's, $n_{dust}(b)\,db$, resembles that of comets (Hanner et al. 1992, Harker et al. 2002, Li \& Greenberg 1998) which we approximate as: 
\begin{equation}
n_{dust}\;=n_{C}\,b^{-4}\,db  
\end{equation}
We use 
 $b_{min}$ and $b_{max}$ to denote the minimum and maximum size, respectively, of the dust from the KBO's while $n_{C}$ is a normalization constant which does not matter for our purposes here.  We may write that the total mass in the dust, $M_{total}$ is
\begin{equation}
M_{total}\;=\;{\int}^{b_{max}}_{b_{min}}\left(\frac{4{\pi}{\rho}_{dust}b^{3}}{3}\right)\,n_{C}b^{-4}\,db
\end{equation}
Similarly, if $M_{orbit}$ denotes the mass of the dust which is large enough
to remain in orbit after it is produced, then
\begin{equation}
M_{orbit}\;=\;{\int}^{b_{max}}_{b_{crit}}\left(\frac{4{\pi}{\rho}_{dust}b^{3}}{3}\right)\,n_{C}b^{-4}\,db
\end{equation}
Since 0.5 of the mass of the KBO's is dust,  we write that
\begin{equation}
f_{orbit}\;=\;0.5\,\frac{M_{orbit}}{M_{total}}
\end{equation}
Therefore, we find from equations (34) - (36) that:
\begin{equation}
f_{orbit}\;=\;0.5\frac{ln\,\left(\frac{b_{crit}}{b_{min}}\right)}{ln\,\left(\frac{b_{max}}{b_{min}}\right)}
\end{equation}

While the values of $b_{min}$ and $b_{max}$ are only poorly known, it seems
for many comets that  $b_{min}$ ${\sim}$ 0.1 ${\mu}$m and $b_{max}$ ${\sim}$ 1 cm
(see McDonnell et al. 1987, Hanner et al. 1992).   Results from meteorites (Kyte \& Wasson 1986, Love \& Brownlee 1995) and the Long Duration Exposure Facility (Zolensky et al. 1995) also show that many particles larger than 100 ${\mu}$m exist in the Solar system.   With these extrapolations from the Solar System,  
we find from equation (37) that  $f_{orbit}$ ${\approx}$ 0.25.    
 
\subsection{Infrared Excess}

We now  elaborate upon the schematic discussion given in $\S$2 to show that the infrared excess around a red giant, $L_{excess}$, directly depends upon the rate of dust production from the KBO's, ${\dot M_{KBO}}$.   Here, 
 we include two additional physical processes not discussed in $\S$2 that control the grain's orbital evolution: 
 stellar wind drag and grain-grain collisions.

The stellar wind drag is straightforward to incorporate into the model.       
Re-writing the notation of Gustafson (1994) and Burns et al. (1979),  the increase by ``drag" in the inward drift velocity over that produced by the Poynting Robertson effect
 is given approximately by the factor (1 +  ${\dot M_{wind}}c^{2}/L_{*}$)
where ${\dot M_{wind}}$ is the stellar wind mass loss rate.  In the Solar
System, this ``drag" created by the anisotropic recoil of solar wind particles off an orbiting dust grain 
typically scales as 0.3 of the Poynting Robertson drag  (see, for example,  Gustafson 1994).  Below, we argue that  the stellar wind
drag  around red giants is of similar magnitude.    
   
Consider now the orbital evolution of  dust particles released from  warmed-up KBO's.  
We assume that as seen from the host star, the  dust ejected from the sublimating KBO's is confined to a solid angle, ${\Omega}_{KBO}$.  If the dust cloud is optically thin,  the excess luminosity scales
directly as the optical depth of the dust.  If $n_{dust,i}$ denotes the density of
dust particles of radius $b_{i}$,  we write that: 
\begin{equation}
L_{excess}\;=\;\left(\frac{{\Omega}_{KBO}}{4{\pi}}\right)\,L_{*}\;{\sum}_{i}\;{\int}_{D_{final}}^{D_{init}}{\pi}b_{i}^{2}\,n_{dust,i}(D)\,dD
\end{equation} 
In equation (38), we assume that all the dust particles are formed
at an initial outer distance from the star, $D_{init}$, and they are all destroyed
at some final, inner radius, $D_{final}$. We also assume that the cross
section of the grains is simply given by their projected surface area, ${\pi}b_{i}^{2}$.  

We now show how the Poynting-Robertson and stellar wind drags
affect the dust motion.
From the equation of continuity, we write:
\begin{equation}
n_{dust,i}\;=\;\frac{\dot N_{dust,i}}{{\Omega}_{KBO}\,D^{2}\,V_{i}(D)}
\end{equation}
where ${\dot N_{dust,i}}$ denotes the rate of production by number of dust grains from KBO sublimation  and $V_{i}(D)$ denotes the inward radial drift speed of the same grains.   If ${\dot M_{dust,i}}$ denotes the rate of production by mass of the i'th type of particle, then: 
\begin{equation}
{\dot N_{dust,i}}\;=\;\frac{3\,\dot M_{dust,i}}{4{\pi}b_{i}^{3}{\rho}_{dust}}
\end{equation}
Since the dust particles drift inwards under the action of the Poynting-Robertson drag, we write: 
\begin{equation}
V_{i}(D)\;=\;\left(\frac{3\,L_{*}}{8\,{\pi}\,b_{i}\,{\rho}_{dust}\,c^{2}\,D}\right)\,\left(1\,+\,\frac{\dot M_{wind}c^{2}}{L_{*}}\right)
\end{equation}  
(Burns et al. 1979, Gustafson 1996).  Note  that since $V$ = $dD/dt$ and if we ignore the correction for stellar  wind drag, equation (1) can be derived from equation (41).  

We are now able to compute the infrared excess from the system.  
Combining equations (38)-(41), we find:
\begin{equation}
L_{excess}\;=\;\frac{1}{2}\;{\sum}_{i}{\dot M_{dust,i}}c^{2}\left({\int}_{D_{final}}^{D_{init}}\frac{dD}{D}\right)\,\left(1\,+\,\frac{\dot M_{wind}c^{2}}{L_{*}}\right)^{-1}
\end{equation}
From $\S$3.6, we write that 
\begin{equation}
{\sum}_{i}\,{\dot M_{dust,i}}\;=\;{\dot M_{KBO}}\,f_{orbit}
\end{equation}
Therefore, evaluating equation (42) and using equation (43): 
\begin{equation}
L_{excess}\;=\;\frac{1}{2}\,{\dot M_{KBO}}\,c^{2}\,f_{orbit}\;\left(1\,+\,\frac{\dot M_{wind}c^{2}}{L_{*}}\right)^{-1}\;ln\left(\frac{D_{init}}{D_{final}}\right)
\end{equation}
Equation (44) is a fundamental result in this paper and shows that the excess infrared luminosity scales directly as the dust production rate, ${\dot M_{KBO}}$. 

It should be recognized that equation (44) is a more
complex version of equation (4) derived in $\S$2.  In equation (44), we employ
${\dot M_{KBO}}$ instead of $M_{KBO}(0)/t_{RGB}$.  Also, in equation (44) we include the wind contribution to limiting the particle's orbital lifetime
in addition to the Poynting-Robertson effect.  Finally, there are  numerical
constants that are different between equations (4) and (44).   

We now estimate  the  parameters required to use equation (44).  First, we  estimate the magnitude of the stellar wind drag.  Currently, 
there are only a few first ascent red giants where the mass loss rate is measured with much confidence.   Robinson, Carpenter \& Brown (1998) report $L_{*}$ and ${\dot M_{wind}}$ for ${\alpha}$ Tau (K5 III) and ${\gamma}$ Dra (K5 III) of 394 L$_{\odot}$ and ${\sim}$1.3 ${\times}$ 10$^{-11}$ M$_{\odot}$ yr$^{-1}$ and 535 L$_{\odot}$ and  ${\sim}$1.0 ${\times}$ 10$^{-11}$ M$_{\odot}$ yr$^{-1}$, respectively.  These results yield values of ${\dot M_{wind}}c^{2}/L_{*}$ of 0.49 and 0.28, respectively.  For HR 6902 (G2 II), the luminosity is 550 L$_{\odot}$ and ${\dot M_{wind}}$ is somewhere between 0.8 and 3.4 ${\times}$ 10$^{-11}$ M$_{\odot}$ yr$^{-1}$ (Kirsch, Baade \& Reimers 2001).
For this star, the correction to the Poynting Robertson drag is between 0.2 and 0.9.  Here,  we simply assume that all red giants have the same stellar wind drag and adopt  
${\dot M_{wind}}c^{2}/L_{*}$ = 0.4, a value for this ratio close to that in the Solar System.  

To use equation (44), we also need to estimate  $D_{final}/D_{init}$.  In the absence of any other effects, the refractory dust would drift inwards until it reached a temperature of ${\sim}$1000 K when it would be destroyed by sublimation.  From equation (16) we can  show that if $T$ ${\sim}$ 170 K at $D_{init}$, then $D_{final}$ ${\sim}$ 0.03 $D_{init}$.  However, before they sublimate, the particles' orbits  might be disrupted as they drift inwards by grain-grain collisions. In such collisions, the grains may shatter into pieces, some of which are small enough
to be driven away from the star by radiation pressure.  Shattering may also
produce more smaller grains with a net increase in surface area of the particles.      
If the radial optical depth through the circumstellar dust is ${\tau}$, then,
depending upon the velocity dispersion of the grains and their size and spatial distributions, the probability during one orbital period, $P$, of a grain-grain collision may  also be of order ${\tau}$.
Therefore, the mean time between collisions is $P/{\tau}$.  With an orbital
period at 45 AU of 300 yr and ${\tau}$ = 10$^{-4}$, the optical depth of interest here since in this case 
the infrared excess becomes detectable,  the mean time between collisions
of 3 ${\times}$ 10$^{6}$ yr, which is comparable to the Poynting Robertson decay
time for a dust particle of radius 100 ${\mu}$m orbiting around a red giant
with a luminosity of 300 L$_{\odot}$. Therefore, as a very simple
first approximation, grain-grain collisions are assumed to  occur when the dust particle's
orbital decay from the Poynting-Robertson drift has significantly changed
its distance from the star.   
 Although detailed calculations are required (see, for example, Wyatt et al. 1999, Krivov, Mann \& Krivova 2000),  we adopt  $D_{final}$ = 0.5 $D_{init}$as a crude representation of the effects of grain-grain collisions. Our results for the infrared excess luminosity are not strongly sensitive to this
value for $D_{init}/D_{final}$ since, from equation (44), L$_{excess}$ depends only logarithmically upon this ratio.   

Using the calculations for ${\dot M_{KBO}}$ described in $\S$3.5 and equation (44), we show in Figure 3 the results 
for $L_{excess}/L_{*}$ as a function of $L_{*}$ for several models.   
We see that  $L_{excess}/L_{*}$ increases until it reaches
its maximum  value which occurs when ${\Delta}a$ ${\approx}$ $a_{break}$.  Models with initially some small KBO's and $a_{break}$ = 1 km show a higher value of ${\dot M_{KBO}}$ than the model with only large KBO's and $a_{break}$ = 200 km.
Once the star's luminosity is high enough that ${\dot M_{KBO}}$ achieves saturation, then 
$L_{excess}$ is approximately constant and $L_{excess}/L_{*}$ decreases directly with $L_{*}$.
  From equations (30) and (44), the  value for L$_{excess}$ when ${\dot M_{KBO}}$ reaches saturation is: 
\begin{equation}
L_{excess}\;=\;\left(\frac{f_{orbit}\,ln\left(\frac{D_{init}}{D_{final}}\right)}{8\,ln\left(\frac{a_{max}}{a_{break}}\right)}\right)\left(\frac{T_{Subl}}{T}\right)\left(1\,+\,\frac{\dot M_{wind}c^{2}}{L_{*}}\right)^{-1}
\left(\frac{M_{large}(0)c^{2}}{t_{RGB}}\right)
\end{equation}
For the standard parameters adopted above, equation (45), an
elaboration of equation (4) derived in $\S$2, yields $L_{excess}$ ${\approx}$ 0.02 L$_{\odot}$ when $T$ = 170 K.

\section{COMPARISON WITH OBSERVATIONS}

In $\S$3, we have calculated the excess luminosity produced by a cloud of sublimating KBO's.  We now show that this radiation may be detectable. 

 To measure the amount of excess radiation that a red giant exhibits at
25 ${\mu}$m, we need to determine its photospheric emission at this wavelength.  Previous studies of first ascent red giants have shown that the bulk of the detected flux at 12 ${\mu}$m, F$_{\nu}$(12 ${\mu}$m), emerges
from a star's photosphere (see, for example, Jura, Webb \& Kahane 2001, Knauer, Ivezic \& Knapp 2001).  Here, we use the usual notation that ${\nu}$ denotes
frequency.  We extrapolate from this measurement of the total flux at 12 ${\mu}$m to estimate the photospheric contribution to F$_{\nu}$(25 ${\mu}$m).   From  IRAS data, Jura (1999) found   for the photospheres of the very brightest red giants that on average, $F_{\nu}$(25 ${\mu}$m)/$F_{\nu}$(12 ${\mu}$m) = 0.233.  We define $f_{ex}(25)$, the excess flux at 25 ${\mu}$m, by the expression:
\begin{equation}
f_{ex}(25)\;=\;\frac{1}{0.233}\,\frac{F_{\nu}(25 {\mu}m)}{F_{\nu}(12 {\mu}m)}\,-\,1
\end{equation}
In Figure 4, we show a histogram of  $f_{ex}(25)$ for 66 nearby red giants identified by a procedure described below.  
Over 90\% of  the red giants exhibit $|f_{ex}(25)|$ ${\leq}$ 0.10  implying that the IRAS data can be used to predict the photospheric flux at 25 ${\mu}$m to within
10\%.

 The histogram shown in Figure 4 is not symmetric around
$f_{ex}(25)$ = 0.  The sample of 66 stars from which the data are obtained
has a mixture of surface gravities and abundances, and the intrinsic distribution of $f_{ex}(25)$ around its mean value may be skewed.  The skewed histogram might also result if some of these 66 stars possess an excess at 25 ${\mu}$m of ${\sim}$ 5\% of the photospheric flux.  Better data should allow us
to confirm the reality of this hint of infrared excesses.

The sample of stars is chosen from 
  nearby  red giants which are least subject to interstellar extinction. We  select stars from the    
 the Yale Bright Star Catalog with
B-V ${\geq}$ 1.0 mag which also lie within 150 pc of the Sun (from {\it Hipparcos} data), have $M_{V}$ ${\leq}$ 0.0 mag and have
200 $L_{\odot}$ $<$ L $<$ 300 $<$ L$_{\odot}$.   We use 200 L$_{\odot}$ as the lower bound of the
luminosities that we consider because, as shown in Figure 3, for stellar luminosities lower than this value, the amount of the infrared excess becomes very sensitive to
the initial numbers of KBO's smaller in radius than 200 km.  We use 300 L$_{\odot}$ as the upper bound for the stellar luminosity because for luminosities higher than this value,
equation (44) begins to overestimate the infrared excess as ${\Delta}a$ approaches $a_{max}$.  We compute the star's luminosity by
assuming no reddening and applying the bolometric corrections from Flower (1996) for red giants with the assumption that $M_{Bol}$ (Sun) = 4.74 mag (Bessell, Castelli \& Plez 1998). 
There are 66 stars which satisfy these criteria and
which were detected with IRAS.  

In our models described in $\S$3, $D_{final}$ = 0.5 $D_{init}$.  Because this change in orbital radius is relatively small, we assume that the grains emit at a single temperature,
determined from equation (16), by their initial distance from the star.  Assuming that the infrared excess is characterized by a 170 K black body,
then $L_{excess}$ ${\approx}$ 1.5 ${\nu}L_{\nu}(25)$ where $L_{\nu}(25)$ denotes
the specific luminosity at 25 ${\mu}$m.  

In Figure 3, in addition to showing the calculations for the standard model of $L_{excess}/L_{*}$ vs. $L_{*}$, we also plot the data for the 66 identified red giants.   For each star, we use its distance and  the non-color corrected flux  from the IRAS Faint Source Catalog or, when that is not available, the flux from the Point Source Catalog, to estimate $L_{\nu}(25)$.   For the stars with $f_{ex}(25)$ $<$ 0.1,  we plot with a horizontal bar  the quantity (0.1) (1.5)${\nu}$L$_{\nu}$(25 ${\mu}$m)/L$_{*}$ vs. L$_{*}$.  For stars with $f_{ex}(25)$ $>$ 0.10, we plot with a filled triangle the quantity $f_{ex}$(25)(1.5)${\nu}$L$_{\nu}$(25 ${\mu}$m)/L$_{*}$ vs. L$_{*}$.    
Of the six stars with $f_{ex}(25)$ $>$ 0.1,  there are three stars, HR 736, HR 7956 and HR 9029, which  also have 60 ${\mu}$m excesses.  The 60 ${\mu}$m excess for HR 9029 also has been noted by Plets et al. (1997).   These infrared excesses are so small that it is very difficult to infer a dust
temperature, and  this evidence is at best a weak hint
that these stars possess sublimating KBO's.  

From Figure 3,  the we see  that the data points usually lie below the theoretical curve for the regime where initially there are  many
KBO's smaller than 200 km in radius. Therefore,  we find that the production of dust from systems containing
comet-like bodies can produce a detectable infrared excess if the total mass of 
the KBO's is at least 0.1 M$_{\oplus}$.
We can also see from Figure 3 that if initially there are only KBO's with a radius
larger than 200 km, then a KBO system similar to that in own Solar System would not
produce a detectable infrared excess around most red giants.  However, this calculation 
was performed with $M_{big}(0)$ =  0.02 M$_{\oplus}$.  If, instead, we assume that  $M_{big}(0)$ ${\approx}$ 0.1 M$_{\oplus}$, then the predicted fluxes would be a factor of five larger than shown. In this case, the data points would then lie below the computed line and an infrared excess would be detectable.   Thus, as anticipated in our discussion in $\S$2, regardless of the exact size distribution,  we expect that an infrared excess might be
detectable if the total mass of the KBO system lying at 45 AU orbital radius is larger than 0.1 M$_{\oplus}$.          

\section{DISCUSSION}
Our main argument is that as a star evolves on the red giant phase, it
is possible to constrain the mass of the KBO's of the system.
Here, we consider how our results compare with other investigations of the outer
regions of our Solar System and its analogs among other stars.  

While sublimation of KBO's at an orbital radius of 45 AU might lead to a detectable infrared excess during a star's red giant evolution,   
our analysis does not provide  strong
constraints on the mass of KBO's much further than 50 AU.
  Melnick
et al. (2001), Ford \& Neufeld (2001) and Ford et al. (2003) have proposed that in order to account for the detected circumstellar H$_{2}$O and OH  around the mass-losing carbon star IRC+10216, there is a cloud of  comets  at ${\sim}$300 AU with a mass of at least ${\sim}$10 M$_{\oplus}$.  In contrast,    our typical upper limit to the mass in comet-like objects is ${\sim}$0.1 M$_{\oplus}$ at ${\sim}$ 45 AU.  This apparent mass gradient of KBO's is remarkable and deserves further study. 

Main sequence stars often possess from 0.01 M$_{\oplus}$ to 0.1 M$_{\oplus}$ of
dust at orbital separations of typically between 10 AU and 100 AU (see Zuckerman 2001).  Our
results for evolving red giants are consistent with the hypothesis that there may be as much as  an additional ${\sim}$ 0.1 M$_{\oplus}$ in icy comets closer than ${\sim}$50 AU around most main sequence stars. However, it appears that macroscopic KBO's with radii greater than 1 km do not
overwhelmingly dominate the mass budgets of solid material  orbiting beyond 10 AU around those main sequence stars with measured amounts of dust.   

The total mass of the KBO's in orbit around our Sun is somewhere between 0.01 M$_{\oplus}$ and 0.1 M$_{\oplus}$.  Even the larger value is much less mass
than predicted by extrapolations from the inner Solar System for the initial
Solar Nebula (Luu \& Jewitt 2002).  Our upper limits for the mass of systems of KBO's around red giants 
shows that  the analogs to the Kuiper Belt also have similarly ``low" masses at orbital radii less than 50 AU.

Very precise  measurements of  the fluxes from red giants in comparison with model atmospheres
may allow for a more reliable determination of whether they
possess infrared excesses.  If the photospheric flux could be estimated
to within 1\%, instead of  10\%, it would be possible to improve  estimates of the  masses of the KBO systems by a factor of 10.
With such a capability, we could determine whether the hint in the data of some stars having KBO's of ${\sim}$0.1 M$_{\oplus}$. 

In this paper, we have focused on the production of an infrared excess by
sublimation of comets.  Our calculations do not pertain to
those red giants where an infrared excess with a characteristic dust temperature of ${\sim}$70 K is measured that is produced by ${\sim}$0.1 M$_{\oplus}$ of dust at ${\sim}$200 AU from the star (Jura 1999).  The origin of this dust is uncertain;  at least
in some cases, confusion with interstellar cirrus is possible.
Further work is required to understand the nature of this material.

\section{CONCLUSIONS}
We have computed the evolution of a system of KBO's when a star becomes a
red giant. For  KBO's at 45 AU from the host star, we find that the dust released during  the sublimation of ice from these
objects might produce a detectable infrared excess around red giants
of luminosities in the range 200 L$_{\odot}$ to 300 L$_{\odot}$ if the mass of the KBO's is at least 0.1 M$_{\oplus}$.   To date, there is no strong evidence for such sublimating KBO's, but with improved data, it
may be possible to detect  systems which are analogs to our Solar System's KBO's around first ascent red giants.

This work has been partly supported by NASA.

\newpage
\begin{figure}
\epsscale{1}
\plotone{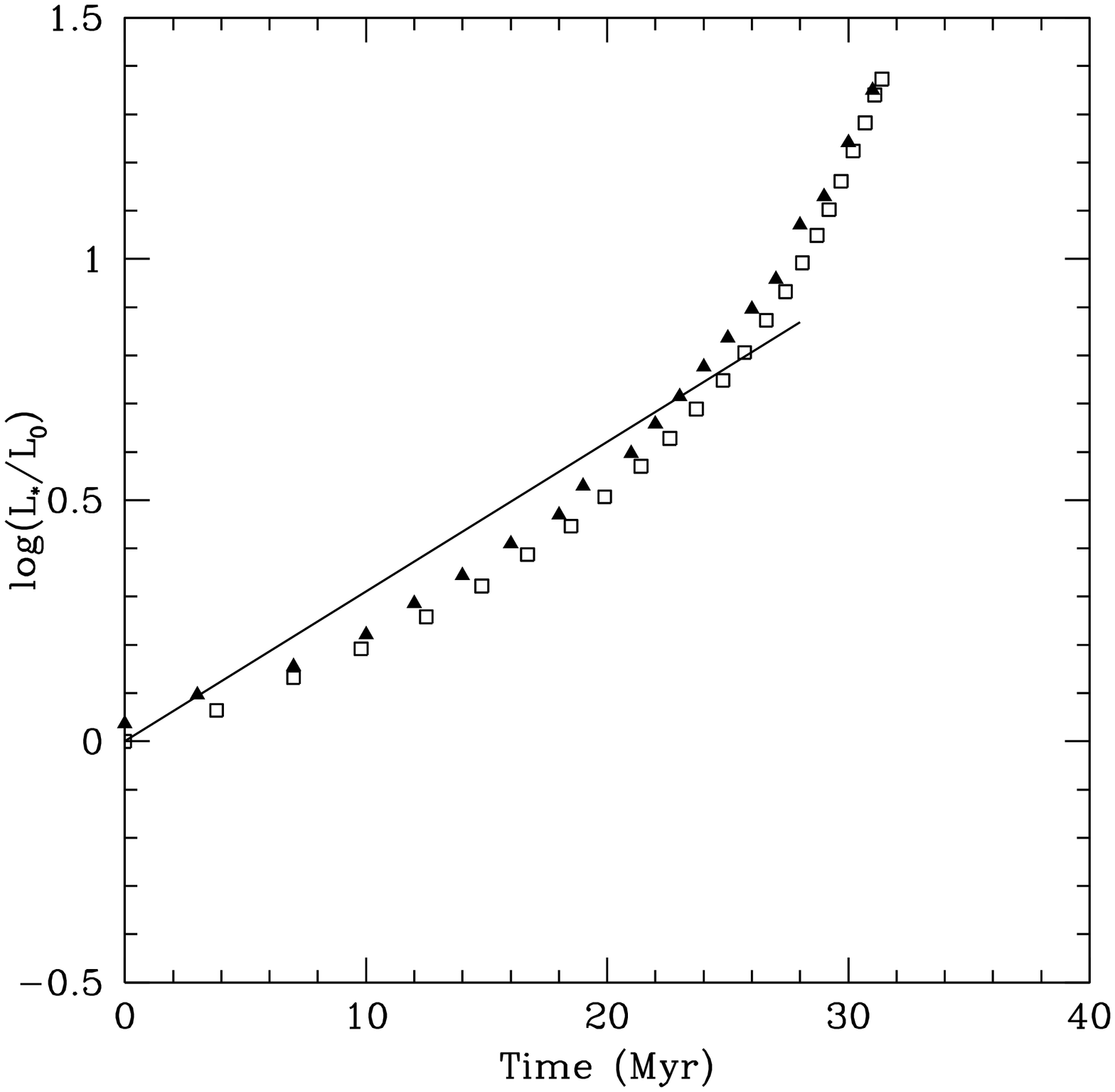}
\caption {We show a plot of theoretical calculations from Girardi et al. (2000) for the luminosity on the Red Giant Branch vs. time.  The filled triangles show the results  for a star  of 1 M$_{\odot}$, $L_{0}$ = 87 L$_{\odot}$ and a starting time of 1.220 ${\times}$ 10$^{10}$ yr while the open squares show
 the calculations  for a star of 1.5 M$_{\odot}$, $L_{0}$ = 91 L$_{\odot}$ and a starting time of 2.7742 ${\times}$ 10$^{9}$ yr. The solid line shows the prediction from equation (12) with $t_{RGB}$ = 14 Myr.  The solid curve lies within 20\% of the detailed calculations for $L$ $<$ 1000 L$_{\odot}$ for both sets of models.}
\end{figure}
\newpage
\begin{figure}
\epsscale{1}
\plotone{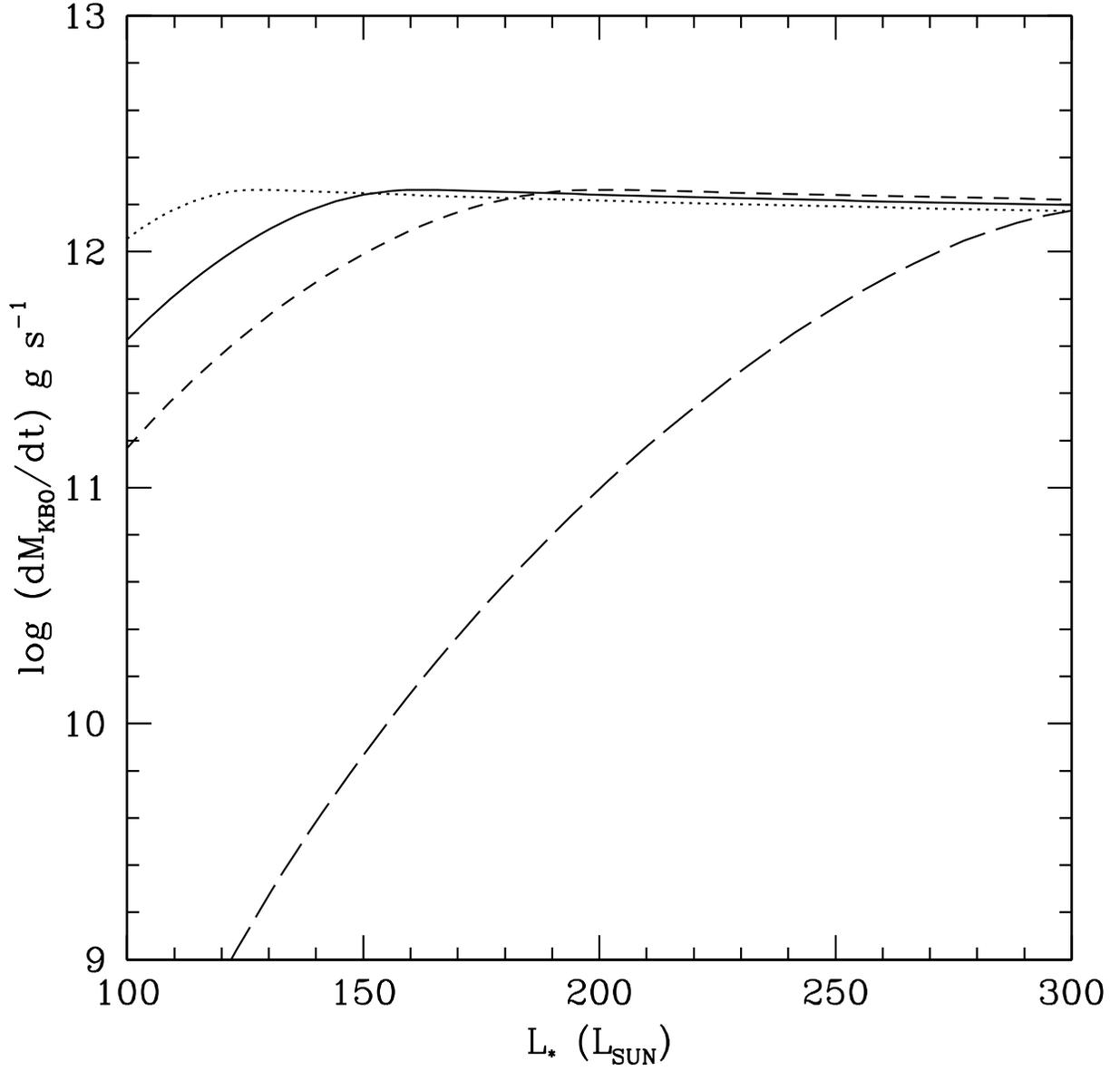}
\caption {Plots of  ${\dot M_{KBO}}$ from a cloud of KBO's vs. the star's luminosity.  The solid line shows the calculations for the standard model consisting of both small and large KBO's with $D_{init}$ = 45 AU and M$_{large}(0)$ = 0.1 M$_{\oplus}$ while the dotted and short-dashed lines show the results for the cases
where $D_{init}$ = 40 AU and 50 AU, respectively. Note the approach to saturation for ${\dot M_{KBO}}$ given by equation (30). The long-dashed line shows the
calculations from equation (31) for a cloud of only big KBO's with M$_{big}(0)$ = 0.02 M$_{\oplus}$ at $D_{init}$ = 45 AU.}
\end{figure}
\newpage
\begin{figure}
\epsscale{1}
\plotone{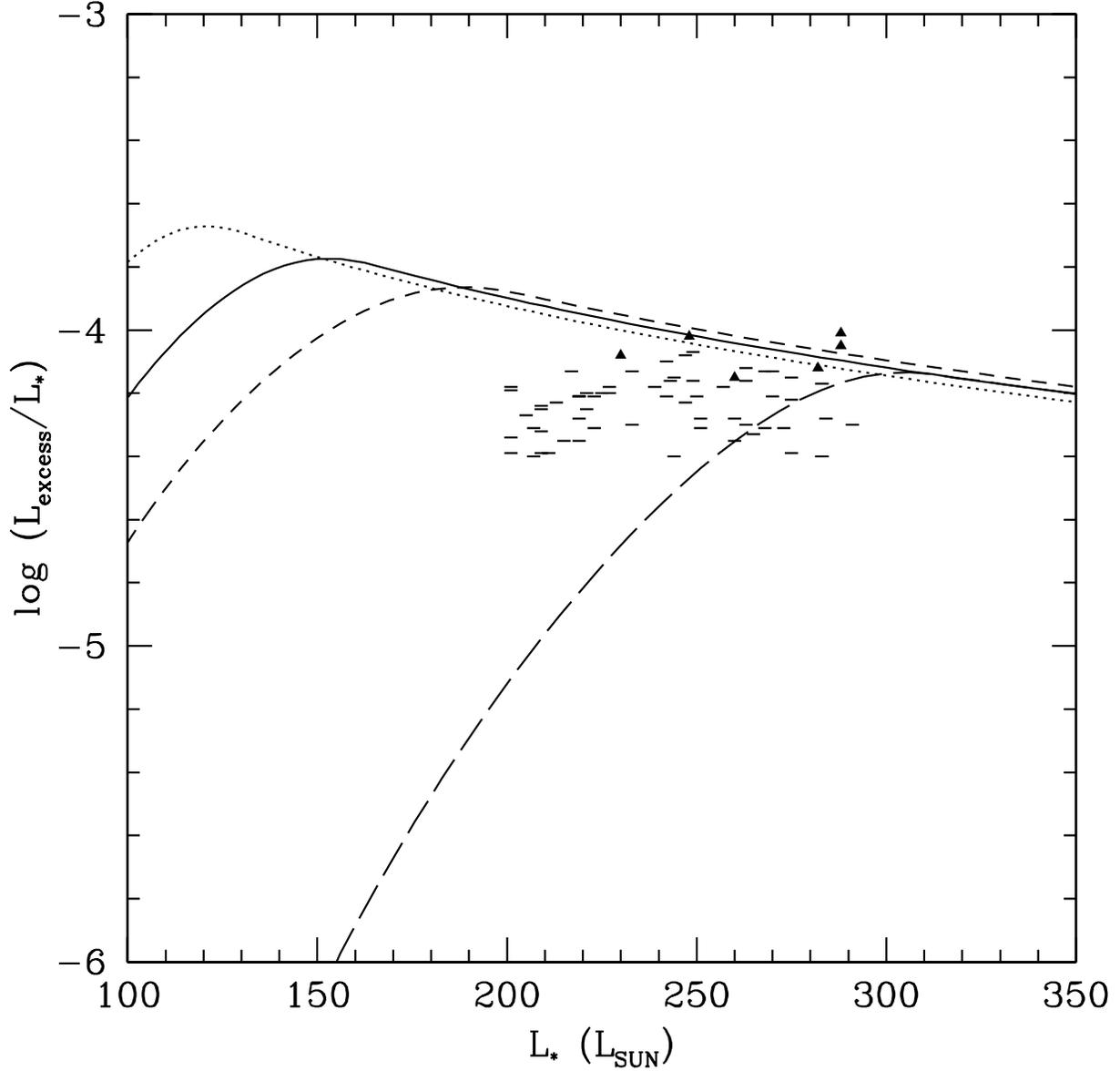}
\caption {The solid curve shows a plot of $L_{excess}/L_{*}$ vs. $L_{*}$ for the standard model with both small and large KBO's described in the text with $D_{init}$ = 45 AU and M$_{large}(0)$ = 0.1 M$_{\oplus}$.  The dotted and dashed lines show calculations for
the same parameters as the standard model except that we use $D_{init}$ = 40 AU and 50 AU, respectively.  The long-dashed line shows the results for the case
with $D_{init}$ = 45 AU and only large KBO's with M$_{big}(0)$ = 0.02 M$_{\oplus}$.   We also display data for 66 stars identified by a procedure described in the text.  The bars represent stars where $f_{ex}(25)$ defined by equation (46) is less than 0.1 while the triangles represent those stars where $f_{ex}(25)$ $>$ 0.1.}
\end{figure}
\newpage
\begin{figure}
\epsscale{1}
\plotone{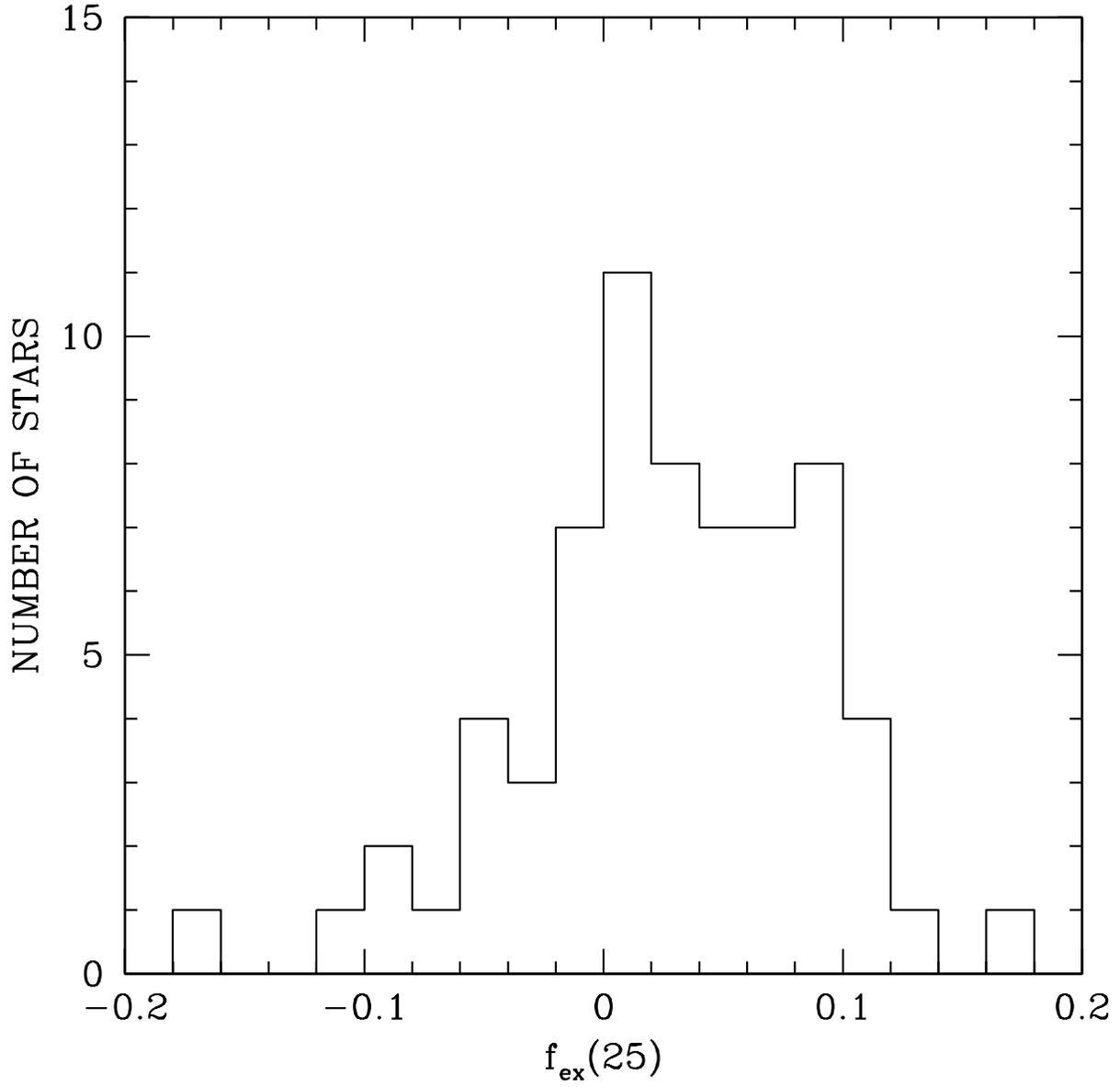}
\caption {A histogram of the fractional excess at 25 ${\mu}$m, $f_{ex}(25)$, defined in equation (35), for the 
stars represented in Figure 3.}
\end{figure}\end{document}